\documentclass[dvips,10pt,cmp,aps,twocolumn,nofootinbib]{revtex4}

\usepackage{amsmath,amssymb,epsf,epsfig,amsthm,bm,graphicx,subfigure}

\begin{document}

\title{Shock waves and Birkhoff's theorem in Lovelock gravity}%

\author{E. Gravanis}
\email{eliasgravanis@netscape.net}
%\affiliation{****}
%\author{S. Willison}
%\email{steve@cecs.cl}
%\affiliation{Centro de Estudios Cientificos}
\date{\today}%

\begin{abstract}
Spherically symmetric shock waves are shown to exist in Lovelock
gravity. They amount to a change of branch of the spherically
symmetric solutions across a null hypersurface. The implications of
their existence for the status of Birkhoff's theorem in the theory
is discussed.
\end{abstract}

\maketitle

\section{introduction}

In general relativity Birkhoff's theorem is roughly the statement
that outside a spherically symmetric (even time-varying) source the
metric field is necessarily static and it is given by the
Schwarzschild
solution~\cite{birkhoff}\cite{Jebsen}\cite{Alexandrow}. A better
statement of the theorem is that the vacuum field equations imply
that a spherically symmetric $C^2$ solution is locally equivalent to
a maximally extended Schwarzschild metric
\cite{Hawking-Ellis-Book}\footnote{Useful additional background on
Birhoff's theorem in Einstein gravity is provided in the
Refs.~\cite{Goenner}\cite{Schmidt:1997mq}.}. An important refinement
came with the work of Ref.~\cite{Bergmann}, building on the work of
Refs.~\cite{Papapetrou}\cite{Papapetrou0}: Schwarzschild metric is
the unique spherically symmetric family of vacuum solutions and
Birkhoff's theorem holds even if we lower differentiability class to
$C^0$. The result which encapsulates, one may say, the essence of
what makes the theorem possible, is that spherically symmetric shock
waves do not exist. In more words, there cannot be a vacuum solution
respecting spherical symmetry whose first derivative becomes
discontinuous across a null hypersurface. This result was obtained
already in \cite{Papapetrou0}. Had such a solution existed, any
statement of Birkhoff's theorem for smooth metrics would be strongly
weakened. On the other hand, the in-existence of the spherically
symmetric shock waves may be attributed to the uniqueness of the
spherically symmetric smooth metric of a given mass. There simply
cannot be a non-trivial matching across a (null) hypersurface
respecting the symmetry.

In dimensions higher than four, Einstein field equations can be
formally thought of as a special case of the Lovelock field
equations \cite{Lovelock:1971yv}. The additional terms are of higher
order in curvature nonetheless they still are second order
differential equations for the metric tensor. Originally, at least
in the recent years, an interest in Lovelock gravity came through
studies on low energy effective actions in string theory
\cite{Zwiebach:1985uq}. In the last decade the interest was revived
due the higher dimensions becoming popular, essentially through
works such as \cite{Antoniadis:1990ew}\cite{Randall:1999vf}. In the
last few years a certain amount of interest in Lovelock gravity was
further drawn, taking also an attractive turn. The shear
viscosity/entropy density bounds for a conformal field theory living
on the boundary of AdS~\cite{Policastro:2001yc}\cite{Kovtun:2003wp}
are modified and restricted when applying causality conditions on
the conformal field theory with a Lovelock gravity dual in the
bulk~\cite{Brigante:2007nu}\cite{Brigante:2008gz}. Moreover these
restrictions coincide unexpectedly to ones obtained through positive
energy conditions on scattering processes in super-conformal field
theories~\cite{Hofman:2008ar}. These findings led to a series of new
works, see e.g.
\cite{Brustein:2008cg,Cai:2008ph,Ge:2009ac,Brustein:2009rn,Shu:2009ax,Cai:2009zn,deBoer:2009pn,Camanho:2009vw,Buchel:2009sk,deBoer:2009gx,Camanho:2009hu,Brustein:2010jb}
related to Lovelock gravity.

Establishing Birkhoff's theorem in Lovelock gravity requires a more
careful phrasing than in general relativity. Perhaps the most
complete presentation of the theorem has been given in
Ref.~\cite{Zegers:2005vx} building on the work of
\cite{Charmousis:2002rc}. Comments in that direction have been
presented in other works
\cite{Deser:2003up,Deser:2005gr,Maeda:2007uu,Bogdanos:2009pc}. The
statement of the theorem is one of uniqueness: The vacuum
spherically symmetric metrics of differentiability class $C^2$ and
for generic values of the couplings of the theory, are locally
equivalent to a specific family of solutions, some of which are the
Lovelock black hole metrics. A characteristic feature of Lovelock
gravity is that these solutions are multi-valued.

Now, what about metrics of differentiability lower than $C^2$? In
Ref.~\cite{Garraffo:2007fi} spherically symmetric \emph{vacuum}
solutions with $C^0$ piecewise $C^\infty$ metrics where explicitly
constructed. Their metric is everywhere smooth except at certain
time-like as well as space-like hypersurfaces where it is only
continuous; its first derivative is discontinuous. That is,
non-trivial \emph{vacuum shells} with time- and space-like
trajectories were explicitly shown to exist. These imply two things.
One, there is a problem of non-causal evolution due to the spacelike
vacuum discontinuities. Second, any staticity interpretation of a
Birkhoff's theorem in this theory is very much weakened by the
possibility of a series of spherical vacuum shells whizzing along
spacetime. The existence of spherical shock waves in Lovelock
gravity, that is of spherical vacuum shells moving along light-like
trajectories, are the subject of the present work. They elegantly
appear as a mere \emph{change of branch} along a null hypersurface.
Now even if one finds a way, by some nice principle, to censor out
the non-null discontinuities, the null ones cannot be excluded. They
are gravitational shock waves, perfectly existent in general
relativity, with the exception of the spherical ones which gives
rise to Birkhoff's theorem. In Lovelock gravity the spherical ones
exist and appear in fact as fairly natural objects in the theory
along with its multiple branch solutions. Presumably, all these
non-trivial vacuum shell spherical symmetric solutions can be
regarded as large (non-linear) scalar perturbations of a given
smooth spherically symmetric metric.

In order to show that configurations such as the ones discussed in
the previous paragraph are indeed solutions of a given theory,
requires to formulate and solve, one way or another, junction or
matching conditions in that theory. In general relativity, the
matching conditions for the time- or space-like hypersurface where
given their final form in Ref.~\cite{Israel:1966rt}. Shock waves and
null hypersurface matching conditions have been studied in various
old and more recent works
\cite{Papapetrou}\cite{Dautcourt,Lichnerowicz,PenroseNull,Taub:1973,Taub:1980zr,RedmountPTP.73.1401,Clarke&Dray,Berezin:1987bc,Barrabes:1991ng}.
The problem we set ourselves to solve is to construct certain shock
wave solutions in Lovelock gravity. To do this we shall exploit the
formulation of matching conditions first presented in
Ref.~\cite{Gravanis:2003aq} and further elaborated in
\cite{Gravanis:2004kx} and \cite{Gravanis:2009pz}. The basic facts
of the method can be summarized as follows. The gravitational field
is described by the vielbein and the spin connection. Consider a
sequence of smooth configurations which are arbitrarily close
approximations of a given discontinuous one, e.g. one where the spin
connection becomes discontinuous across hypersurfaces [which may
also intersect]. The hypersurfaces divide spacetime up into `bulk'
regions where fields are smooth. At the hypersurfaces the
discontinuous fields are ill-defined. Then the action functional of
the theory evaluated on that sequence can be shown to converge to a
new action functional involving only well defined information:
fields outside the hypersurfaces. The fields in the vicinity of each
side of the hypersurface contain all the information about the
discontinuity, therefore nothing more needs to appear in the
equations. [That presumably means that no explicit reference to the
induced fields on the hypersurface is required.] Working this way
the hypersurfaces need only be locally smooth; their causal
character, null or non-null, is irrelevant. The whole analysis is
done off-shell, therefore the equations of motion for the
discontinuous fields obtained by Euler-Lagrange variation of the new
action are the well defined limit of the usual equations of motion
for the smooth fields. The new action contains explicitly terms for
each hypersurface. The matching conditions are the parts of the
equations of motion with support at the hypersurfaces. The
convergence of the off-shell action to the new action makes them
well defined: the bulk fields on the sides of a hypersurface of
\emph{any} smooth configuration approximating well the discontinuity
must obey those matching conditions.

The way of `matching' in the formulation of
\cite{Gravanis:2003aq,Gravanis:2004kx,Gravanis:2009pz} is somewhat
different than usual: One works with globally defined fields with a
certain amount of discontinuity at some places and attempts to write
down an action functional for these fields which is well defined in
the sense stated above. The spirit of the analysis reflects nicely
the old work of Papapetrou and Treded~\cite{Papapetrou} in general
relativity we already mentioned. The principle is that field
configurations of differentiability lower than the order of the
field equations, are also solutions of the theory if they can be
approximated arbitrarily well by sequences of smooth
\emph{solutions}. A subtlety here is that such sequences is not
always easy to make explicit. One such case is the vacuum solutions
we are interested in. On the other hand the well-defined-ness of the
matching condition is established by sequences of smooth
configurations off-shell. There are always non-vacuum smooth
configurations approximating arbitrarily well a given vacuum
discontinuous configuration. In that limiting sense solutions of the
matching conditions may be regarded as meaningful stationary point
of the classical action.

As the formulation of the matching conditions we shall use is not a
well known one, we will devote the section \ref{the lagrangian and
equations of motion} in explaining the technical and physical
details leading to the matching condition, equation
(\ref{junction-theoretical}). In section \ref{Shock waves} the
formulation is suitably applied to obtain the alleged shock wave
solution, and certain additional comments are given in the final
section.

\section{Lovelock gravity}

To fix ideas consider a specific example of Lovelock gravity. Let
our theory be Einstein gravity with cosmological constant $\lambda$
supplemented by the quadratic Lovelock term in five dimensions. This
is usually termed as Einstein-Gauss-Bonnet gravity. We denote the
coupling of the quadratic Lovelock term by $\alpha$. It has length
dimension $L^{+2}$. The Lagrangian of the theory reads
\begin{align}\label{egb tensorial-1}
&S=-\lambda \int_M \sqrt{-g}\,+\\
\frac{1}{2\kappa^2} &  \int_M \sqrt{-g} \left\{R+\alpha (R^2-4
R_{\alpha\beta}R^{\alpha\beta}+R_{\alpha\beta\gamma\delta}R^{\alpha\beta\gamma\delta})\right\}\,.
\nonumber
\end{align}

The spherically symmetric $C^2$ solutions of the theory (\ref{egb
tensorial-1}) are the Boulware-Deser
metrics~\cite{Boulware:1985wk,Wheeler:1985nh,Wheeler:1985qd,Wiltshire:1985us}
(their general forms in Lovelock gravity where first studied in
\cite{Myers:1986un}):
\begin{align}\label{BD metric}
ds^2&=-f_{BD} dt^2+f^{-1}_{BD}dr^2+r^2 d\Omega^2\,, \\
f_{BD}&=1+\frac{r^2}{4\alpha}\left\{1\pm
\sqrt{1+\frac{4\alpha\kappa^2 \lambda}{3}+\frac{16\alpha
M}{r^4}}\right\}\,.\nonumber
\end{align}
$d\Omega^2$ is a metric of the unit round 3-sphere. $M$ is a
constant of integration related to the mass of the solution. The
Boulware-Deser metrics \emph{are not single-valued}: they have two
branches, corresponding to the $\pm$ sign choice, which have very
different properties. For one thing, there are no black hole
solutions in the metrics of the branch corresponding to the sign
$+$. Also for $\lambda=0$ this branch of the solution is not
asymptotically flat: it is asymptotically anti-de Sitter for
$\alpha>0$ and de Sitter for $\alpha<0$. A related fact is that only
the metrics of the branch $-$ converge to Einstein gravity metrics
in the limit $\alpha \to 0$. It has been argued that the asymptotic
vacuum of the branch $+$ metrics is unstable due the ghost
excitations \cite{Boulware:1985wk}\cite{Charmousis:2008ce}. On this
basis one may discard it from the beginning, but that is rather too
quick. These metrics are an inherent property of Lovelock gravity
and their effects should be studied before they are discarded with
reason, or not discarded at all. For one thing, their presence might
make things better or worse, stability- or other-wise, for the well
behaved branch metrics; both situations are of physical importance.
One should bear in mind that the two types of metrics are naturally
intermingled by the theory. Lovelock gravity allows $C^0$ piecewise
smooth non-trivial vacuum configurations to be cut-and-paste
constructed out of metrics of the same or different branch as shown
in Ref.~\cite{Garraffo:2007fi}, see also \cite{Gravanis:2007ei}. New
such configurations associated with shock waves are presented here.

\section{the lagrangian and equations of motion}
\label{the lagrangian and equations of motion}

\subsection{First order formalism}

Proceeding with the mathematical analysis, it is far  more
convenient than working with the metric tensor $g_{\mu\nu}$ directly
to use the vielbein $E^{~a}_{\mu}$ and an $SO(4,1)$ spin connection
$\omega^{~ab}_\mu$ one-forms over $M$ as our variables. They will be
treated as differential forms $E^a \equiv dx^\mu E_{\mu}^{~a}$,
$\omega^{ab}\equiv dx^\mu \omega^{~ab}_\mu$. We will do our
diffentiations and integrations using exterior
calculus\footnote{Convenient references for this formulation are
\cite{Zumino:1985dp} and \cite{Eguchi:1980jx}.}. The metric tensor
$\eta_{ab}=(-+\cdots+)$ and the volume anti-symmetric tensor (form)
$\epsilon_{abcde}$ are the invariant tensors $SO(4,1)$. From them we
can build invariant forms over the spacetime manifold.

We will use a convenient notation for contraction with the volume
form, for example
\begin{equation*}
\epsilon(\Omega E^3):=\epsilon_{abcde} \Omega^{ab} \!\wedge\! E^c
\!\wedge\! E^d \!\wedge\! E^e\,.
\end{equation*}
Its convenience can be seen in variations and differentiations, e.g.
$\delta\{\epsilon(\Omega E^3)\}=\epsilon(\delta \Omega
E^3)+3\,\epsilon(\Omega E^2\, \delta E)$. The invariance of the
contraction also implies that
\begin{equation*}\label{}
d(\epsilon(\cdots))=\epsilon(D(\cdots))\,,
\end{equation*}
where $D$ is the covariant derivative associated with the spin
connection. From now on the wedge symbol $\wedge$ will be dropped in
the wedge product of forms.

The spin connection throughout this work will be Levi-Civita i.e.
torsion-free: $T^a:=DE^a \equiv dE^a+\omega^a_{~b}\, E^b=0$. $d$ is
the exterior calculus derivative operator. Its nilpotence, $dd=0$,
holds on all $C^2$ forms. Then the affine connection on the tangent
bundle is Levi-Civita and defines the usual covariant derivative of
general relativity. The spin connection is not a tensor. The
curvature
$\Omega^a_{~b}:=d\omega^a_{~b}+\omega^a_{~c}\,\omega^c_{~b}$ is an
$SO(4,1)$ tensor and a two-form over $M$. We can write that more
compactly treating the forms $\omega$ and $\Omega$ as matrices:
$\Omega=d\omega+\omega^2$. The curvature satisfies the Bianchi
identity: $D\Omega \equiv d\Omega+\omega\Omega-\Omega \omega=0$
identically, as one may verify.

The curvature form is related to Riemann tensor by the relation
$\Omega^{ab}=\frac{1}{2}E^{~a}_{\mu} E^{~b}_{\nu}
R^{\mu\nu}_{~~\rho\sigma}\,dx^{\rho} dx^{\sigma}$. In other words to
obtain the Riemann tensor form the curvature form one has to invert
the matrix $E^{~a}_{\mu}$. We shall not need to do that as we will
work exclusively with forms, though invertibility will not be evaded
anywhere. Thus the veilbein and connection formulations is
equivalent to the metric formulations. Then the action (\ref{egb
tensorial-1}) can be written in the following form
\begin{equation}\label{egb in forms}
S=-\frac{1}{2}c_0\int_M \epsilon(E^5)+\frac{1}{2}c_1 \int_M
\epsilon(\Omega E^3) +\frac{1}{2}c_2 \int_M \epsilon(\Omega\Omega
E)\,.
\end{equation}
The coupling constants $c_0$, $c_1$ and $c_2$ are related to the
more usual couplings by
\begin{equation}\label{coupling_relations}
c_0\equiv\frac{\lambda}{60}\,, \quad c_1
\equiv\frac{1}{3!\,\kappa^2}\,, \quad
c_2\equiv\frac{\alpha}{\kappa^2}\,,
\end{equation}
and $\kappa^2=8\pi G$, where we introduce $G$ as the Newton's
constant (though conventions differ in the literature). The
cosmological constant appears usually as $\Lambda \equiv \kappa^2
\lambda$. We shall prefer using mostly the $c$ couplings. The length
dimension of $c_1$ is $L^{-3}$ and the dimension of $c_2$ is
$L^{-1}$.

\subsection{Manufacturing a discontinuity}
\label{Manufacturing a discontinuity}

Imagine then spacetime consisting of three regions: two open bulk
regions $N$, $\bar N$ and a closed region $\Delta$ surrounding the
hypersurface $\Sigma$ we would like to be locus of the
discontinuity. The action may written as
\begin{equation*}\label{}
S=\int_{N} {\cal L}+\int_\Delta {\cal L}+\int_{\bar N}{\cal L}\,,
\end{equation*}
where ${\cal L}$ is a Lovelock gravity Lagrangian.

Consider the limit $\Delta \to \Sigma$. It is not that this part of
spacetime actually shrinks but rather the values of the fields $E$
and $\omega$ are moved. This can be done by choosing a sequence of
configurations imitating this process. On the common boundary of
$\Delta$ with each of $N$ and $\bar N$ the values of the fields $E$
and $\omega$ are held fixed (and in general different). These values
become the values of the fields on each side of $\Sigma$ as it is
embedded in each of $N$ and $\bar N$. These are \emph{bulk} fields
and they are well defined as long as there is a coordinate
neighborhood covering the hypersurface $\Sigma$.

The limit we described here as $\Delta \to \Sigma$, or any limit
equivalent to it, was studied for any theory built out of forms
fields alone, such as $E$ and $\omega$, in
Ref.~\cite{Gravanis:2009pz}. The general result is as follows.
Evaluate the action $S$ on an arbitrary sequence of configurations
consistent with the limit $\Delta\to \Sigma$ as described above.
Then for each such sequence of configurations one obtains a sequence
of functionals, whose convergence needs to be checked. Convergence
depends on the amount of discontinuity allowed at $\Sigma$. That
should be formalized into specific conditions on the fields, the
`continuity conditions'. Choosing them appropriately the action
\emph{converges} to a new action, which necessarily depends only on
well defined quantities, the bulk fields. [That convergence we
usually describe as `well-defined-ness' of the new action.] Such
continuity conditions are not unique and depend on the theory; in
general they can be quite strange. In the limit $\Delta \to \Sigma$
the action necessarily involves distributional fields; on the other
hand the new action depends only on the fixed bulk fields on the
sides of $\Sigma$. Everything is done off-shell, therefore whatever
holds for the actions it holds also for the equations of motion. The
convergence of the action to the new action means then two things.
First, the equations of motion of the distributional fields are
meaningful, and secondly, they have been neatly disentangled, the
necessary integrations having been essentially done, into the form
of the equations of motion deriving from the new action. One may
then say that the new action is an equivalent form
of the action of the theory %in the presence of discontinuities. **
for discontinuous fields.

A straightforward application of these ideas is offered by Lovelock
gravity under the most usual continuity condition. In
\cite{Gravanis:2009pz} it was shown that that one may write down a
new action for Lovelock gravity in the presence of discontinuities
if the vielbein $E$ is continuous and the spin connection $\omega$
is discontinuous, a result essentially contained also in
\cite{Gravanis:2004kx}. Lorentz transformations are still a local
symmetry of the new
action~\cite{Gravanis:2003aq}\cite{Gravanis:2004kx}. Therefore our
continuity condition explicitly is: The vielbein is continuous
across the hypersurface modulo a Lorentz transformation. This
continuity condition will be applied and explained in more detail in
section \ref{Shock waves}.

Based on these results we may write the action (\ref{egb in forms})
in the presence of the discontinuity at $\Sigma$ in the equivalent
form:
\begin{align}\label{I total}
    S= &-\frac{1}{2}c_0\int \epsilon(E^5)\\
    & +\frac{1}{2}c_1 \int \epsilon(\Omega
    E^3)+\frac{1}{2}c_1\int_\Sigma\epsilon((\bar\omega-\omega)E^3)\nonumber\\
    &+\frac{1}{2}c_2 \int \epsilon(\Omega\Omega
    E)\nonumber\\
    &+\frac{1}{2}c_2 \int_\Sigma
    \epsilon((\bar\omega-\omega)\{(\bar\Omega+\Omega)-\frac{1}{3}(\bar\omega-\omega)^2\}
    E) \nonumber\,.
\end{align}
The bulk integrals are over $N\bigcup \bar N$.  Writing
$(\bar\omega-\omega)^2$ matrix multiplication of the forms is
understood. $\omega$ and $\Omega$ are respectively the spin
connection and its curvature on $N$ evaluated at $\Sigma$; more
precisely, as $N$ is open, they are the limits of these variables as
we approach $\Sigma$ from within $N$. Similarly for $\bar\omega$ and
$\bar\Omega$ in $\bar N$. The veilbein is continuous modulo a
Lorentz transformation therefore we can express everything in terms
of a single veilbein field $E$.

\subsection{Equations of motion}

Equations of motion are obtained by varying the action with respect
to the fields, $E$ and $\omega$.

It was shown in \cite{Gravanis:2003aq} and more elegantly in
\cite{Gravanis:2004kx} that the variation of the action w.r.t. the
spin connection $\omega$ trivially vanishes upon imposing that
torsion is zero i.e. that the connection provides the familiar
Levi-Civita covariant derivative and (\ref{egb in forms}) is indeed
equivalent to the Lovelock gravity action (\ref{egb tensorial-1}).
That is, all equations of motion, including the matching conditions,
are obtained by varying w.r.t. the vielbein $E$. This is a merely
algebraic variation of the Lagrangian.

Explicitly, equations of motion are obtained for every smooth
submanifold in the problem, that is, the bulk regions, the
hypersurfaces, and in the general case the intersections of the
hypersurfaces. There is a term in the Lagrangian for each one of
these submanifolds. Let $\delta E^a=\lambda(x)^a_{~b}\, E^b$ be the
variation of the vielbein, where $\lambda(x)^a_{~b}$ is an arbitrary
smooth field with support in any one of the submanifolds. The
equations of motion for that submanifold are obtained from the
corresponding term in the Lagrangian. From the bulk regions we
obtain the usual smooth field equations of motion; from the
hypersurfaces and their intersection we obtain a set of matching
conditions. If in general there are other fields with energy tensor
$T^a_{~b}$, the field equations of each submanifold read
\begin{equation}\label{field eq general 1}
\delta {\cal L}=T^a_{~b}\, \lambda(x)^b_{~a}
\end{equation}
times its volume element.

In vacuum, $T^a_{~b}=0$. That is,
\begin{equation}
\delta {\cal L}=0\,.
\end{equation}
In vacuum there is no need to identify components; one simply sets
to zero all non-trivial terms in this equation.

Here, we have the bulk regions $N$, $\bar N$ and the hypersurface
$\Sigma$ to deal with.

The field equations in the region $N$ read
\begin{equation}\label{bulk field equations}
-5c_0\,\epsilon(E^4 \delta E)+3c_1\, \epsilon(\Omega E^2 \delta E)
+c_2\, \epsilon(\Omega\Omega \delta E)=0\,.
\end{equation}
A similar expression holds in $\bar N$ for the barred fields.

Consider the spherically symmetric metric
\begin{equation}\label{spherimetric}
ds^2=-g^2 dt^2+ \frac{dr^2}{g^2}+ r^2 d\Omega^2\,,
\end{equation}
where $g=g(r)$ and $d\Omega^2$ is the metric of the unit round
3-sphere. The metric (\ref{spherimetric}) can be written in terms of
the vielbein one-forms
\begin{equation}\label{5-E}
E^0 =g\,dt\,, \quad E^1=\frac{dr}{g}\,, \quad E^i=r\, \tilde E^i\,.
\end{equation}
$\tilde E^i$ is a veilbein of the unit round 3-sphere. We denote by
$\tilde \omega^{ij}$ its Levi-Civita connection. The curvature of
$\tilde \omega^{ij}$ necessarily is $\tilde \Omega^{ij}= \tilde E^i
\tilde E^j$. The Levi-Civita connection $\omega^{ab}$ of $E^a$ reads
\begin{equation}\label{5-omega}
\omega^{01}=g' g\, dt\,, \quad \omega^{i1}=g\, \tilde E^i\,, \quad
\omega^{ij}=\tilde \omega^{ij}\,,
\end{equation}
The curvature $\Omega^{ab}$ of the connection $\omega^{ab}$ reads
\begin{align}\label{5-Omega}
& \Omega^{01}=-\frac{(g^2)''}{2}E^0 E^1\,, \\
& \Omega^{0i}=-\frac{(g^2)'}{2r}E^0 E^i\,, \quad
\Omega^{1i}=-\frac{(g^2)'}{2r}E^1 E^i\,, \nonumber \\
& \Omega^{ij}=\frac{1-g^2}{r^2} E^i E^j\,. \nonumber
\end{align}
It is then straightforward to show that the entire content of
(\ref{bulk field equations}) amounts to a single differential
equation linear in $g^2-1$. Its solution is
\begin{equation}\label{egb metric 1}
g^2-1=\frac{3c_1}{2c_2}\, r^2 \Big\{1 \pm \sqrt{1+\frac{c_2
\lambda}{27c_1^2}+\frac{C}{r^4}}\Big\}\,.
\end{equation}
Translating the $c$ couplings into the more usual ones via
(\ref{coupling_relations}) we indeed obtain the Boulware-Deser
metric (\ref{BD metric}) with $g^2=f_{BD}$. The integration constant
$C$ is of course related to the mass of the solution.

For future use let us note the following. Denote by $g_{BD\pm}$ the
function $g$ for the respective branch of the Boulware-Deser
solution. One observes that
\begin{equation}\label{plus minus relation}
g^2_{BD+}+g^2_{BD-}=2+\frac{3c_1}{c_2}\, r^2\,.
\end{equation}
This relation will be useful later on.

The vacuum field equation $\delta {\cal L}=0$ for $\Sigma$ is the
matching condition
\begin{align}\label{junction-theoretical}
     &\ i_\Sigma^*\big\{\frac{3}{2}c_1\, \epsilon((\bar\omega-\omega)E^2 \delta E)+\\
    +&\frac{1}{2}c_2\,
    \epsilon((\bar\omega-\omega)\{\bar\Omega+\Omega-\frac{1}{3}(\bar\omega-\omega)^2\}
    \delta E)\big\}=0\,. \nonumber\,
\end{align}
$i^*_\Sigma$ denotes the pull-back of the form into $\Sigma$.

Equation (\ref{junction-theoretical}) is derived under the
hospitable assumption that the vielbein in spacetime is continuous
across $\Sigma$ modulo a Lorentz transformation.

When $\Sigma$ is time- or space-like, there is a unique geometry on
$\Sigma$ induced from the bulk. One may choose a vielbein adapted to
$\Sigma$ [all but one its components are tangentially oriented].
Pulled back into $\Sigma$ defines an induced vielbein, which is
regarded as intrinsic to $\Sigma$. Solving the zero torsion
condition one obtains a unique Levi-Civita spin connection for that
vielbein. This is because projecting $\eta_{ab}$ into those
tangential directions leaves us with an non-degenerate (invertible)
tensor. The obtained connection coincides with the tangential
components of $\omega$. The induced fields from $N$ and $\bar N$ may
differ only by a Lorentz transformation. That is there is indeed a
unique geometry on the hypersurface inherited from the bulk, and is
regarded as intrinsic to it. Equation (\ref{junction-theoretical}),
which in general involves a `matter' energy tensor $T^a_{~b}$ on its
r.h.s. according to (\ref{field eq general 1}), it is equivalent to
the matching conditions first written down in
\cite{Davis:2002gn}\cite{Gravanis:2002wy}. In those works the
matching conditions appeared in the usual kind of formulation (as
set by the work of Israel \cite{Israel:1966rt}) which involves
explicitly the intrinsic geometry of $\Sigma$ and the extrinsic
curvatures of $\Sigma$ w.r.t. the bulk regions $N$ and $\bar N$. The
equivalence of that formulation to (\ref{junction-theoretical}) was
shown in \cite{Gravanis:2003aq} and re-visited in
\cite{Gravanis:2007ei}.

When $\Sigma$ is null, the induced vielbein may also be regarded as
an intrinsic vielbein; it does span the space of tangential
one-forms. The difference is that one direction is null and the
projection of $\eta_{ab}$ into the tangential direction is
degenerate (non-invertible). There is no unique Levi-Civita
connection for the induced vielbein, thus no unique inherited
geometry. Penrose \cite{PenroseNull}, see also \cite{Clarke&Dray},
classifies (three) different induced geometries which can be
regarded as intrinsic to the hypersurface; these geometries differ
on the level the induced notions of parallel transport along
$\Sigma$ agree. These notions, presumably, depend on the derivatives
of the metric. All this wealth of structure available in the null
hypersurface is reflected in practice as follows. The normal vector
is null therefore orthogonal to itself [and indeed tangential on the
hypersurface]. Thus one cannot construct projections along the
tangential directions of $\Sigma$ just by knowing its normal vector.
That, combined with the non-uniqueness of the intrinsic geometry,
implies that splitting the fields into components containing
information intrinsic and extrinsic to $\Sigma$ is less
straightforward in the null case. Formulations of the matching
conditions involving explicitly the intrinsic geometry of the
hypersurface necessarily have qualitative differences in the null
and non-null case. Unifying descriptions of the two cases can be
found~\cite{Clarke&Dray}\cite{Barrabes:1991ng} but they rather
emphasize the peculiarities of the null case.

In our formulation of the matching conditions, leading to
(\ref{junction-theoretical}), only \emph{bulk field} information in
the vicinity of $\Sigma$ is involved. The intrinsic geometry
information exists implicitly in the continuity conditions on the
fields. That is, it exists in a minimal choice [so that to
accommodate as many cases as cases as possible] of which field
variables are continuous across $\Sigma$; the rest are
discontinuous, as restricted of course by the field equations. Thus
the causal character of $\Sigma$, time- or space-like, or null,
enters the matching conditions only through our requirements on the
fields in each specific problem. Formula
(\ref{junction-theoretical}) holds in all cases.

In a most general setting, studied in Ref.~\cite{Gravanis:2009pz},
any chosen continuity conditions is such that the matching
conditions are well defined, in the sense explained previously. On
this basis, the acceptable continuity conditions are certainly not
unique \cite{Gravanis:2009pz}. In practice, one chooses acceptable
continuity conditions which are geometrically intelligible. As we
have already mentioned, our choice here is a usual one: We have
required that the veilbein in spacetime is continuous across
$\Sigma$ modulo a Lorentz transformation, which means that the
metric tensor $g=\eta_{ab} E^a \otimes E^b$ in spacetime is
continuous.

It is worth to mention that the choice of the vielbein $E^a$ and the
connection $\omega^a_{~b}$ as gravitational field variables plays an
important role. A choice of variables involving, for example, the
metric tensor would necessarily involve its inverse. In any such
case our formulation of the problem in terms of globally defined
discontinuous fields would become complicated and almost none of the
tools used in
Refs.~\cite{Gravanis:2003aq,Gravanis:2004kx,Gravanis:2009pz} could
be applied effectively in order to derive equations such as
(\ref{junction-theoretical}) and prove facts about them. Moreover,
the one-forms $E^a$ and $\omega^a_{~b}$ are tensor valued in the
Lorentz group. They can be written w.r.t. to arbitrary coordinates
in \emph{each} region in spacetime. The continuity condition on the
veilbein across their common boundaries i.e. the hypersurfaces, is
that the vielbein is continuous modulo Lorentz transformations, a
local symmetry of the action (\ref{I total}). Indeed, in every
specific problem, one can construct a Lorentz transformation between
the veilbeins on the sides of a hypersurface. This transformation is
then used in (\ref{junction-theoretical}) in order to express all
fields in it w.r.t. the same basis; this completes the statement of
the matching conditions. Therefore there is no need to explicitly
use special coordinates chosen such that continuity is ensured. In
this sense the formulation employed here parallels the work in
Ref.~\cite{Barrabes:1991ng}, though the freedom to use any
convenient coordinates on each side of the hypersurface is achieved
quite differently.

In all, one has in hand a formulation of the matching conditions
which can be uniformly applied to the space- and time-like as well
as to the null hypersurface which is the case of interest. Of course
writing down the correct equations of motion is different than
understanding their content. In the null case the wealth of
structure is revealed only after systematic general analysis. A full
discussion of the null hypersurface matching conditions is far
beyond the scope of the present work. In what follows we apply
(\ref{junction-theoretical}) to show that the shock waves we
advertised at the beginning of our work indeed exist.

\section{Shock waves}
\label{Shock waves}

Continuity of the veilbein in spacetime $M$ means the following.
Each veilbein field, $E$ and $\bar E$, can be extended from the
region it is defined, $N$ and $\bar N$ respectively, across $\Sigma$
and into a neighborhood of the other region. In the overlap $E$ and
$\bar E$ may differ by a Lorentz transformation which is a local
symmetry of action (\ref{I total}). Let us put this condition into
formulas for the problem of interest.

On the manifolds $N$ and $\bar N$ the vielbein is given respectively
by
\begin{equation}\label{E}
E^0 =g\,dt\,, \quad E^1=\frac{dr}{g}\,, \quad E^i=r\, \tilde E^i
\end{equation}
and
\begin{equation}\label{bar E}
\bar E^{\bar 0}=\bar g\, d\bar t\,, \quad \bar E^{\bar
1}=\frac{d\bar r}{\bar g}\,, \quad \bar E^i=\bar r\,\tilde E^i\,.
\end{equation}
Recall that $\tilde E^i$ is a veilbein on the unit round 3-sphere
introduced in the previous section, formulas (\ref{spherimetric})
and (\ref{5-E}).

Let $\Sigma$ be a hypersurface along which
\begin{equation}\label{null trajectories}
\frac{d r}{g}-g\, dt=0\,, \quad  \frac{d \bar r}{\bar g}-\bar g\,
d\bar t=0\,,
\end{equation}
in $N$ and $\bar N$ respectively. More precisely, the pull-back of
the forms on the l.h.s. of these relations into $\Sigma$ vanishes.
$\Sigma$ is null.

A null hypersurface in spacetime dimension five is generated by a
three-parameter family of null geodesics, one through each point of
the hypersurface. The null geodesics along $\Sigma$ satisfy
(\ref{null trajectories}), with parameters the points on the unit
3-sphere. Along these geodesics $ds^2=0$ thus the induced metric is
degenerate. The degenerate induced metric is another way, especially
when working directly with the metric tensor, to recognize a null
hypersurface.

The field $E^a$ is assumed extendible across $\Sigma$ and into the
manifold $\bar N$. Similarly the field $\bar E^{\bar a}$ is assumed
extendible across $\Sigma$ and into the manifold $N$. Explicitly
that means the following. There is a neighborhood of $M$ around
$\Sigma$ where $E^a$ and $\bar E^{\bar a}$ co-exist. This is the
overlap of their support in spacetime. In that neighborhood define
fields $E^{\bar a}$ and $\bar E^a$ such that
\begin{align}\label{1}
& \bar E^{a}\:\: \textrm{is a rotated}\:\: \bar E^{\bar a}\:\:
\textrm{so that}\:\: \bar E^a|_\Sigma=E^{a}|_\Sigma\,, \\
& E^{\bar a}\:\: \textrm{is a rotated}\:\: E^a\:\: \textrm{so
that}\:\: E^{\bar a}|_\Sigma=\bar E^{\bar a}|_\Sigma\,, \nonumber
\end{align}
By `rotated' we of course mean Lorentz transformed
\begin{equation}\label{2}
\bar E^a=\Lambda^a_{~\bar a}\, \bar E^{\bar a}\,,
\end{equation}
everywhere in the overlap of $E^a$ and $\bar E^{\bar a}$ in $M$. The
inverse of the matrix $(\Lambda^a_{~\bar a})$ is the matrix
$(\Lambda^{\bar a}_{~a})$. Therefore $\bar E^{\bar a}=\Lambda^{\bar
a}_{~a}\,\bar  E^a$ in that neighborhood. Completely analogous
formulas hold for the unbarred fields as we extend $E^a$ into $\bar
N$. We shall not need explicitly these formulas.

Continuity of the $a=i$ components of the vielbein implies that
$r=\bar r$ along $\Sigma$. That implies that
\begin{equation}\label{null r}
i^*_\Sigma dr=i^*_\Sigma d \bar r\,.
\end{equation}
In view of (\ref{null trajectories}), $t+\int dr g^{-2}$ is a
natural coordinate along $\Sigma$ in the spacetime region $N$.
Similarly $\bar t+\int d\bar r \bar g^{-2}$ in $\bar N$. Therefore
the forms $dt +g^{-2} dr$ and $d\bar t +\bar g^{-2} d\bar r$ may
agree at $\Sigma$ modulo suitable factors [null vectors have no
natural normalization]. Indeed, setting $g^2 dt+dr=\bar g^2 d\bar
t+d \bar r$ at $\Sigma$ is consistent with (\ref{null r}).
Translating that in vielbein components we write
\begin{equation}\label{22}
g(E^0+E^1)|_{\Sigma}=\bar g(\bar E^{\bar 0}+\bar E^{\bar
1})|_{\Sigma}\,.
\end{equation}

It is then not hard to show that condition (\ref{22}) implies a
Lorentz transformation $\Lambda^a_{~b}$ which reads
\begin{gather}
\Lambda^a_{~\bar a}=
\left(%
\begin{array}{cc}
\dfrac{\bar g^2+g^2}{2\bar g g} \ & \ \dfrac{\bar g^2-g^2}{2\bar g
g}
 \\
& \\
\dfrac{\bar g^2-g^2}{2\bar g g} \ & \
\dfrac{\bar g^2+g^2}{2\bar g g} \\
\end{array}%
\right).
\end{gather}
We suppress the trivial transformation of the angular components.

This transformation had to be explicitly known in order to express
all tensor valued forms in the same basis: the fields $\bar
\omega^{\bar a}_{~\bar b}$ and $\bar \Omega^{\bar a}_{~\bar b}$ must
be transformed to obtain the components of $\bar\omega$ and
$\bar\Omega$ in the directions of the basis $E^a$, which is the
spacetime veilbein field appearing in equation
(\ref{junction-theoretical}).

Under a Lorentz transformation $\Lambda^a_{~\bar a}$ the spin
connection $\bar\omega$ transforms as
\begin{equation}\label{connection transformation}
\bar\omega^a_{~b}=\Lambda^a_{~\bar a}\, \bar \omega^{\bar a}_{~\bar
b}\, \Lambda^{\bar b}_{~b}+\Lambda^a_{~\bar b}\, d\Lambda^{\bar
b}_{~b}\,.
\end{equation}
The transformation of its curvature tensor $\bar\Omega$ follows and
reads
\begin{equation}\label{}
\bar\Omega^a_{~b}=\Lambda^a_{~\bar a}\, \bar \Omega^{\bar a}_{~\bar
b}\, \Lambda^{\bar b}_{~b}\,.
\end{equation}
These transformations hold in the overlap of the support of $E$ and
$\bar E$ in $M$. Under (\ref{2}) and (\ref{connection
transformation}) the torsion tensor of $\bar E$ transforms as $d\bar
E^{ a}+\bar \omega^{a}_{~b} \bar E^{ b}=\Lambda^a_{~\bar a}(d\bar
E^{\bar a}+\bar \omega^{\bar a}_{~\bar b} \bar E^{\bar b})$. Being
indeed a tensor, its components in both bases vanish
simulataneously. Therefore $\bar\omega^a_{~b}$, given by
(\ref{connection transformation}), are indeed the components of the
Levi-Civita connection $\bar\omega$ in the basis which coincides
with $E^a$ at $\Sigma$. Everything needed in order to apply
(\ref{junction-theoretical}) has now been made explicit.

The quantities of importance are the pull-backs of the `jumps'
$(\bar\omega-\omega)^a_{~b}$ into $\Sigma$. Recalling
(\ref{5-omega}) a straightforward calculation gives
\begin{align}\label{jumps}
& i^*_\Sigma(\bar \omega-\omega)^0_{~1}=0\,,\\
& i^*_\Sigma(\bar \omega-\omega)^i_{~1}=+\frac{\bar
g^2-g^2}{2g}\tilde E^i\,, \nonumber \\
& i^*_\Sigma(\bar \omega-\omega)^i_{~0}=-\frac{\bar
g^2-g^2}{2g}\tilde E^i\,. \nonumber
\end{align}
The jump of the angular components $\omega^i_{~j}=\tilde
\omega^i_{~j}$ vanishes trivially. It is then easy to calculate the
pull-back into $\Sigma$ of the jump squared $((\bar
\omega-\omega)^2)^a_{~b}=(\bar \omega-\omega)^a_{~c} (\bar
\omega-\omega)^{b}_{~c}$, which vanishes identically. [Such a
simplification, due to the high symmetry of the configuration, was
vaguely conjectured in Ref.~\cite{Barrabes:2005ar}.] Using
(\ref{5-Omega}) the forms $(\bar \Omega+\Omega)^a_{~b}$ are also
straightforwardly calculated.

Substituting everything in the matching condition
(\ref{junction-theoretical}) one finally finds
\begin{align}\label{matching conditions explicit}
&-\Big\{3c_1+c_2\frac{2-g^2-\bar g^2}{r^2}\Big\}\, \frac{\bar
g^2-g^2}{2gr}\, \times \\
& \times\, i^*_\Sigma\, \epsilon_{01ijk} E^i E^j E^k \delta
E^-=0\nonumber
\end{align}
where $E^-:=E^0-E^1$. Also $E^+:=E^0+E^1$. The vielbein $(E^\pm,
E^i)$ is a basis adapted to $\Sigma$.

$E^-$ is the normal to $\Sigma$ component of the vielbein in the
sense that $i^*_\Sigma E^-=0$, by (\ref{null trajectories}). The
presence of $\delta E^-$ in (\ref{matching conditions explicit}) is
a reminiscent of the tangent nature of the null normal vector.
Explicitly the variation reads
\begin{equation}
i^*_\Sigma \delta E^-=\lambda(x)^-_{~+}\, i^*_\Sigma E^++
\lambda(x)^-_{~j}\, i^*_\Sigma E^j\,,
\end{equation}
for an arbitrary smooth field $\lambda(x)^a_{~b}$ with support on
$\Sigma$. The component of the matching condition (\ref{matching
conditions explicit}) involving $\lambda(x)^-_{~+}$ it is not
trivial. Therefore (\ref{matching conditions explicit}) implies that
\begin{align}\label{matching conditions explicit 2}
\Big\{3c_1+c_2\frac{2-g^2-\bar g^2}{r^2}\Big\}\, \frac{\bar
g^2-g^2}{2gr}=0\
\end{align}
for all $r$ along $\Sigma$. The solution $\bar g^2=g^2$ amounts to
no discontinuity at all. The non-trivial solution is
\begin{align}\label{matching conditions explicit 3}
3c_1+c_2\frac{2-g^2-\bar g^2}{r^2}=0\,.
\end{align}
Recalling (\ref{plus minus relation}) we see that this equation can
indeed be satisfied, and in fact $g$ and $\bar g$ must belong to
\emph{different branches} of the \emph{same} Boulware-Deser solution
given in (\ref{BD metric}). Thus we obtain the branch changing shock
wave solution we claimed to exist in the beginning of this paper.

\section{comments}

Mathematically, the spherical symmetric shock wave solutions of
Lovelock gravity have been rather established. These shock waves are
possible due to the multi-valued-ness of the spherically symmetric
metrics of the theory. On the other hand, the multi-valued-ness
itself is a symptom of a deeper pathology.

The whole thing can be better understood considering a simpler
system. An elementary such system is a point particle with
Lagrangian $L=\mu v^2+gv^4$. $v$ is the velocity of the particle and
$\mu$ and $g$ are non-zero constants. The canonical momentum reads
$p=2\mu v+4 g v^3$. Varying the action $\int dt L$ one obtains the
Euler-Lagrange equation, $\dot p=0$, the dot denoting time
derivative. That is, $p$ is constant in time. Let across some
instant of time momentum change from $p_\textsl{in}$ to
$p_\textsl{out}$ and velocity from $v_\textsl{in}$ to
$v_\textsl{out}$. The equation of motion requires that
$p_\textsl{in}=p_\textsl{out}$. This translates to
\begin{equation}\label{factorization}
(v_\textsl{in}-v_\textsl{out})(2\mu+4g(v_\textsl{in}^2+v_\textsl{in}
v_\textsl{out}+v_\textsl{out}^2))=0\,.
\end{equation}
This is a matching condition. Along with the solution
$v_\textsl{in}-v_\textsl{out}=0$ i.e. velocity is constant, one
obtains a solution in which velocity jumps according to the
condition $v_\textsl{in}^2+v_\textsl{in}
v_\textsl{out}+v_\textsl{out}^2=-\mu/(2g)$. The jumps may happen at
arbitrary times. We learn that a source-free motion is not
necessarily a uniform motion in this system. In fact, $p=
\textrm{constant}$ may correspond to an infinity of different
piece-wise uniform motions.

From the Hamiltonian point of view, the problem arises in the
following form. The Hamiltonian is a function of the canonical
momentum $p$. Solving $p=2\mu v+4 g v^3$ for $v$ the answer is
multi-valued. Each one of the multiple solutions $v(p)$ defines a
different `canonical branch' of the theory, as we may call it, with
a different Hamiltonian $H=p\, v(p)-L$. We have seen that the
source-free motion $p=\textrm{constant}$ is neither simple nor
unique in this theory, and in general involves sudden changes of
velocity according to the equation $v_\textsl{in}^2+v_\textsl{in}
v_\textsl{out}+v_\textsl{out}^2=-\mu/(2g)$. These jumps are a sudden
change of \emph{canonical} branch as $v_\textsl{in}$ and
$v_\textsl{out}$ are values of the velocity given by different
$v(p)$. Thus the Hamiltonian by which the system evolves may change
abruptly without a cause. A byproduct of this fact is that energy is
not conserved in the free motion of this system. Moreover, one
cannot uniquely relate the space of solutions to the space of the
initial data
% no well defined infinitesimal evolution
i.e. there is no notion of classical phase space for this theory.

All that translates as it is in Lovelock gravity. The role of
velocity is played by the extrinsic curvature which jumps across a
hypersurface in vacuum. The Lagrangian is not linear in the
curvature i.e. not quadratic in the extrinsic curvature. This leads,
in the quise of many components and complicated formulas, to the
same problems encountered in the simple point particle system. The
multi-valued-ness of the Lovelock gravity Hamiltonian and some of
its implications have been emphasized in Ref.~\cite{TZ}. In
Ref.~\cite{Garraffo:2007fi} it was shown by explicit examples that
appropriate jumps of the extrinsic curvature across non-null
hypersurfaces are allowed in vacuum, i.e. the canonical momentum
does not change across the hypersurface nonetheless the theory
changes canonical branch. Mathematically the results arise through
factorized expressions analogous to equation (\ref{factorization}):
There is a continuity imposing factor which we require not to
vanish, and another factor originating from the higher order terms
which leads to a \emph{soluble} equation. That was realized here in
the form of equation (\ref{matching conditions explicit 2}) leading
to equation (\ref{matching conditions explicit 3}) and our shock
wave solutions. The factorization can be seen already at the level
of equation (\ref{junction-theoretical}) which is a complicate
analogue of (\ref{factorization}). In fact factoring out $\bar
\omega-\omega$ in (\ref{junction-theoretical}) one obtains a rank
three tensor whose vanishing is the general condition that must be
satisfied by the fields in a change of canonical branch. The
multiple canonical branches are due to the non-linearity of the
Lagrangian in curvature, an outcome of which is the
multi-valued-ness of solutions.

The shock waves presented here are spherically symmetric and exist
in dimension higher than four. It takes the peculiar dynamics of
Lovelock gravity for them to arise. Now it is worth to mention that,
even in Einstein gravity, once in higher dimensions spherical
symmetry is not a necessary ingredient of Birkhoff's type of
theorems. The reason why is that Israel's black hole uniqueness
theorem \cite{Israel:1967wq}\cite{Israel:1967za}, which says that
static and asymptotically flat black holes are necessarily
spherically symmetric, is not as strong in higher dimensions as it
is in four: If one drops asymptotic flatness then the geometry of
the `angular' manifold [with the topology of the sphere] is not
necessarily that of the round sphere, see the discussion
in~\cite{Gibbons:2002bh}\cite{Gibbons:2002th}. Einstein field
equations restrict the `angular' manifold geometry only that much.
Now all that `room to spare' might make possible even time-dependent
solutions as it is numerically argued in Ref.~\cite{Bizon:2005cp}
using an angular manifold with the topology of the three-sphere but
not the geometry of the round sphere. In fact all that freedom is
controlled once one switches on the Lovelock gravity terms available
in those higher dimensions, as emphasized in the
Refs.~\cite{Dotti:2005rc}\cite{Bogdanos:2009pc}. In this work we
have seen that in Lovelock gravity $C^0$ piecewise smooth
time-dependent solutions exist in the form of shock waves even under
spherical symmetry. The derivation goes through as it is with minor
modifications in any other case of interest [one simply uses an
`angular' vielbein $\tilde E^i$ with the appropriate properties]
e.g. the Boulware-Deser-Cai metrics~\cite{Cai:2001dz} or the metrics
of Ref.~\cite{Bogdanos:2009pc} solutions of theory (\ref{egb
tensorial-1}). Formulas become only more cumbersome, not essentially
different, when including cubic or higher Lovelock terms in the
appropriate dimension. The analogues of the shock waves presented
here arise as long as the smooth solutions are multi-valued through
elementary identities such as relation (\ref{plus minus relation}).
\\\\

\acknowledgments The author would like to thank Steve Willison for
helpful comments.

\bibliography{semi}
\bibliographystyle{h-physrev5}

\end{document}